# Translation of Nagumo's Foundational Work on Barrier Functions: On the Location of Integral Curves of Ordinary Differential Equations*

*original title: „Über die lage der integralkurven gewöhnlicher differentialgleichungen“

Marcel Menner
Aurora Flight Sciences (A Boeing Company)
Cambridge, MA, USA
menner@ieee.org

Eugene Lavretsky
The Boeing Company
Huntington Beach, CA, USA
eugene.lavretsky@boeing.com

*Abstract*— **In 1942, Prof. Mitio Nagumo published his seminal paper on the location of integral curves of ordinary differential equations. Nagumo's paper provides the foundation of the set invariance of ordinary differential equations and barrier functions, which have recently gained popularity for the control design of safety critical dynamical systems. This translation shall serve the community with an easily accessible version of the original 1942 paper in English. A copy of Nagumo's paper in German is also attached as a reference. That copy was created by the Boeing Company, Germany, in an attempt to improve pdf format readability of the original paper.**

## Preface and Glossary

In order to preserve the spirit of Nagumo's original work [1], this translation makes only minor changes to the notation with the purpose of increasing readability. For example, this translation uses time $t$ to emphasize the paper's relevance for dynamical systems. Further, some notation used in the original paper is archaic and/or was more common in the German literature.

The following changes are made:

| | | |
|---|---|---|
| $t$ | (original: $x$) | = time |
| $\boldsymbol{x}$ | (original: $\boldsymbol{y}$) | = state of differential equation |
| $\mathfrak{D} \cap \mathfrak{E}$ | (original: $\mathfrak{D} \cdot \mathfrak{E}$) | = set operator for intersection |
| lim sup | (original: $\overline{\lim}$) | = limit of supremum |
| lim inf | (original: $\underline{\lim}$) | = limit of infimum |
| $[a, b]$ | (original: $\langle a, b \rangle$) | = closed interval |
| $\{x_1, \dots, x_k\}$ | (original: $[x_1, \dots, x_k]$) | = discrete set |
| $\leq, \geq$ | (original: $\leqq, \geqq$) | = less/greater than or equal to |

Nagumo uses Dini derivatives [2] defined as

$$\overline{D}_+ f(t) := \limsup_{h \to +0} \frac{f(t+h) - f(t)}{h}, \quad \overline{D}_- f(t) := \limsup_{h \to -0} \frac{f(t+h) - f(t)}{h},$$

$$\underline{D}_+ f(t) := \liminf_{h \to +0} \frac{f(t+h) - f(t)}{h}, \quad \underline{D}_- f(t) := \liminf_{h \to -0} \frac{f(t+h) - f(t)}{h},$$

where the signs $D_+$ and $D_-$ indicate the limit being approached “from the right” and “from the left”, respectively.

# On the Location of Integral Curves of Ordinary Differential Equations

By Mitio Nagumo

## I. Introduction

In this note, $k$-dimensional vectors are denoted by bold letters. Therefore, under the equation

$$\frac{d\boldsymbol{x}}{dt} = \boldsymbol{f}(t, \boldsymbol{x}) \tag{1}$$

we understand a system of differential equations

$$\frac{dx_i}{dt} = f_i\,(t, x_1, \dots, x_k)$$
$$(i = 1, 2, \dots, k).$$

O. Perron [3] has given a proof of the existence of solutions to an ordinary differential equation $\frac{dx}{dt} = F(t, x)$ in the form that they are enclosed in a region $a \leq t < b$, $\omega_1(t) \leq x \leq \omega_2(t)$, where $\omega_i(t)$ satisfies the conditions

$$D_{\pm}\omega_1(t) \leq F\big(t, \omega_1(t)\big), \qquad D_{\pm}\omega_2(t) \geq F\big(t, \omega_2(t)\big).$$

This aspect has been extensively developed by Hukuhara [4]. He also refers to a subset $\mathfrak{E}$ of the point set $\mathfrak{D}$ in the $(t, \boldsymbol{x})$-space as "majorant to the right in $\mathfrak{D}$" if every solution curve of (1) lying in $\mathfrak{D}$ and starting from any initial point $(t_0, \boldsymbol{x}_0)$ in $\mathfrak{E}$ remains in $\mathfrak{E}$ for $t \geq t_0$.

The main goal of this note is to provide necessary and sufficient conditions for $\mathfrak{E}$ to be a majorant to the right in $\mathfrak{D}$. The conditions are given using lower integrals[1], which are based on the idea of Okamura [5] used for the uniqueness condition of the solution of (1).

## II. Admissible Set

**Definition 1.** *A set of points $\mathfrak{M}$ in the $(t, \boldsymbol{x})$-plane is called admissible to the right for the differential equation (1) if, for every point $(t_0, \boldsymbol{x}_0)$ in $\mathfrak{M}$, there exists a positive number $l$ such that there exists an integral curve of (1) lying in $\mathfrak{M}$ with the initial point $(t_0, \boldsymbol{x}_0)$ for $t_0 \leq t < t_0 + l$.*

Similarly, admissibility to the left can be defined analogously.

**Theorem 1**: Let $\mathfrak{D}$ be an open set in the $(t, \boldsymbol{x})$-plane and let $\mathfrak{E}$ be a closed set in $\mathfrak{D}$, wherein $\boldsymbol{f}(t, \boldsymbol{x})$ is continuous. Then, $\mathfrak{E}$ is admissible to the right for (1) if, and only if, for every point $(t_0, \boldsymbol{x}_0)$ in $\mathfrak{E}$ and any positive $\varepsilon$, there exists a point $(t_1, \boldsymbol{x}_1)$ in $\mathfrak{E}$ satisfying the conditions: $t_0 < t_1 < t_0 + \varepsilon$ and

$$\left|\frac{\boldsymbol{x}_1 - \boldsymbol{x}_0}{t_1 - t_0} - \boldsymbol{f}(t_0, \boldsymbol{x}_0)\right| < \varepsilon.$$

**Proof**. Only the sufficiency of the condition needs to be proven because the necessity is clear. For a point $(t_0, \boldsymbol{x}_0)$ in $\mathfrak{E}$, there exist positive numbers $l$ and $M$ such that the region $\mathfrak{D}_1$: $t_0 \leq t \leq t_0 + l$, $|\boldsymbol{x} - \boldsymbol{x}_0| \leq (M + 1)l$ lies in $\mathfrak{D}$, and $|\boldsymbol{f}(t, \boldsymbol{x})| \leq M$ in $\mathfrak{D}_1$. Let $P_0, P_1, P_2, \dots, P_n = (t_n, \boldsymbol{x}_n)$ be a sequence of points satisfying the conditions: $t_{\nu-1} < t_\nu < t_{\nu-1} + \varepsilon$, $P_\nu \in \mathfrak{E}$, and

[1] The definition will be provided in Section IV.

$$\left|\frac{\boldsymbol{x}_v - \boldsymbol{x}_{v-1}}{t_v - t_{v-1}} - \boldsymbol{f}(t_{v-1}, \boldsymbol{x}_{v-1})\right| < \varepsilon \qquad (\varepsilon < 1). \tag{2}$$

Let $\mathfrak{M}$ be the set of all possible points $P_n$ satisfying these conditions. The upper limit $\xi$ of the values of $t$ for which $(t, \boldsymbol{x}) \in \mathfrak{M}$ satisfies $\xi > t_0 + l$. If this were not the case, $\mathfrak{M}$ would be contained in $\mathfrak{D}_1$. There exists a limit point $(\xi, \boldsymbol{x}^*)$ of $\mathfrak{M}$ on $t = \xi$. Since $(\xi, \boldsymbol{x}^*) \in \mathfrak{E}$, there exists a $(\xi_1, \boldsymbol{x}_1^*) \in \mathfrak{E}$ such that $\xi < \xi_1 < \xi + \varepsilon$ and

$$\left|\frac{\boldsymbol{x}_1^* - \boldsymbol{x}^*}{\xi_1 - \xi} - \boldsymbol{f}(\xi, \boldsymbol{x}^*)\right| < \varepsilon. \tag{3}$$

However, there exists a finite sequence $P_0, P_1, P_2, \dots, P_n$ satisfying the conditions $t_{v-1} < t_v < t_{v-1} + \varepsilon$ and (2) in $\mathfrak{M}$, such that $P_n$ approaches arbitrarily close to $(\xi, \boldsymbol{x}^*)$. Therefore, $t_n < \xi_1 < t_n + \varepsilon$ and according to (3),

$$\left|\frac{\boldsymbol{x}_1^* - \boldsymbol{x}_n}{\xi_1 - t_n} - \boldsymbol{f}(t_n, \boldsymbol{x}_n)\right| < \varepsilon.$$

Consequently, $P_{n+1} = (\xi_1, \boldsymbol{x}_1^*) \in \mathfrak{M}$ with $\xi_1 = t_{n+1} > \xi$, contradicting the definition of $\xi$.

Now let $\boldsymbol{x} = \boldsymbol{X}_N(t)$ be the equation of the curve connecting a sequence $P_0, P_1, P_2, \dots, P_n$ satisfying the conditions $P_v \in \mathfrak{E}$, $t_{v-1} < t_v < t_{v-1} + \varepsilon$, and (2), where $\varepsilon = \frac{1}{N}$. The curves $\boldsymbol{x} = \boldsymbol{X}_N(x)$ lie in $\mathfrak{D}_1$ for $t_0 \le t \le t_0 + l$ and satisfy the inequality

$$|\boldsymbol{X}_N(t') - \boldsymbol{X}_N(t'')| \le (M+1)|t' - t''|.$$

There is then a subsequence $\{N_i\}$ of natural numbers such that $\boldsymbol{X}_{N_i}(t)$ converges uniformly to a continuous curve $\boldsymbol{x} = \boldsymbol{X}(t)$ in $[t_0, t_0 + l]$ as $N_i \to \infty$, which lies in $\mathfrak{E}$. It can be easily proven that for sufficiently large $N_i$,

$$\left|\frac{\boldsymbol{X}_{N_i}(t') - \boldsymbol{X}_{N_i}(t)}{t' - t} - \boldsymbol{f}\big(t, \boldsymbol{X}(t)\big)\right| < \varepsilon$$

if $|t' - t| < \delta, t_0 \le t \le t_0 + l$, where $\delta > 0$ is sufficiently small. From this, it follows that for $N_i \to \infty$ and then for $\delta \to 0$,

$$\frac{d}{dt}\boldsymbol{X}(t) = \boldsymbol{f}(t, \boldsymbol{X}(t))$$

for $t_0 \le t < t_0 + l$ and $\boldsymbol{X}(t_0) = \boldsymbol{x}_0$. QED.

**Theorem 2**. Let $\mathfrak{D}$ and $\mathfrak{E}$ be as in Theorem 1. If $\mathfrak{E}$ is admissible to the right for (1), then every integral curve of (1) lying in $\mathfrak{E}$ can be extended up to the boundary of $\mathfrak{D}$.

**Proof**. Left to the reader.

As an application of Theorem 1 and Theorem 2, we obtain the following result without much difficulty: Let $\boldsymbol{f}(t, \boldsymbol{x})$ be continuous in the range $a \le t < b, \omega_1(t) \le x \le \omega_2(t)$, where $\omega_i(t)$ are continuous in $a \le t < b$ and satisfy the relations

$$\underline{\boldsymbol{D}}_+\omega_1(t) \le f\big(t, \omega_1(t)\big), \qquad \overline{\boldsymbol{D}}_+\omega_2(t) \ge f(t, \omega_2(t)).$$

Then there exists at least one continuous solution $x = x(t)$ of $\frac{dx}{dt} = f(t, x)$ in $a \le t < b$ with the conditions $x(a) = x_0, (\omega_1(a) \le x_0 \le \omega_2(a))$, and

$$\omega_1(t) \le x(t) \le \omega_2(t)$$

for $a \le t < b$.

## III. Operation $\overline{\boldsymbol{D}}_{+[f]}\boldsymbol{\varphi}$

**Definition 2**. A function $\varphi(t, \boldsymbol{x})$ defined on a set $\mathfrak{D}$ in the $(t, \boldsymbol{x})$-space is said to belong to the class $(L)$, more precisely to the class $(L, \alpha)$, in $\mathfrak{D}$, if it is continuous in $\mathfrak{D}$ and there exists a constant $\alpha$ such that for any $(t, \boldsymbol{x}_i) \in \mathfrak{D}$ $(i = 1, 2)$ with a common value of $t$, the inequality holds:

$$|\varphi(t, \boldsymbol{x}_1) - \varphi(t, \boldsymbol{x}_2)| \le \alpha|\boldsymbol{x}_1 - \boldsymbol{x}_2|.$$

If $\varphi(x, \boldsymbol{y})$ is a real-valued function of class $(L)$ on $\mathfrak{D}$, then the limit

$$\limsup_{h \to +0} \frac{\varphi(t_0 + h, \boldsymbol{x}(t_0 + h)) - \varphi(t_0, \boldsymbol{x}_0)}{h}$$

always has the same value, as long as $\boldsymbol{x}(t)$ is any function such that $(t, \boldsymbol{x}(t)) \in \mathfrak{D}$ for $t_0 < t < t_0 + \delta$[2] and

$$\lim_{h\to+0} \frac{\boldsymbol{x}(t_0+h)-\boldsymbol{x}_0}{h} = \boldsymbol{f}(t, \boldsymbol{x_0}).$$

We denote this limit as $\boldsymbol{D}_{+[f]}\varphi(t_0, \boldsymbol{x}_0)$.

Therefore,

$$\overline{\boldsymbol{D}}_{+[f]}\varphi(t_0, \boldsymbol{x}_0) = \limsup_{h\to+0} \frac{\varphi\,(t_0+h, \boldsymbol{x_0}+h\boldsymbol{f_0}) - \varphi\,(t_0, \boldsymbol{x_0})}{h}$$

as long as $(t_0 + h, \boldsymbol{x_0} + h\boldsymbol{f_0}) \in \mathfrak{D}$ for sufficiently small $h \geq 0$, where $\boldsymbol{f_0} = \boldsymbol{f}(t_0, \boldsymbol{x_0})$.

The following relations can be easily proven for functions of class $(L)$:

$$\overline{\boldsymbol{D}}_{+[f]}[\varphi_1(t,\boldsymbol{x}) + \varphi_2(t,\boldsymbol{x})] \leq \overline{\boldsymbol{D}}_{+[f]}\varphi_1(t,\boldsymbol{x}) + \overline{\boldsymbol{D}}_{+[f]}\varphi_2(t,\boldsymbol{x})$$

$$\overline{\boldsymbol{D}}_{+[f]}[\varphi_1(t,\boldsymbol{x}) \cdot \varphi_2(t,\boldsymbol{x})] \leq \varphi_1(t,\boldsymbol{x}) \cdot \overline{\boldsymbol{D}}_{+[f]}\varphi_2(t,\boldsymbol{x}) + \varphi_2(t,\boldsymbol{x}) \cdot \overline{\boldsymbol{D}}_{+[f]}\varphi_1(t,\boldsymbol{x})$$

if $\varphi_i(t, \boldsymbol{x}) \geq 0$ $(i = 1, 2)$.

The following theorem can be easily proven:

**Theorem 3**. Let $\mathfrak{D}$ be a set on which $f(t, \boldsymbol{x})$ is continuous, and let $\varphi(t, \boldsymbol{x})$ be a function of class $(L)$ on $\mathfrak{D}$. Let $\mathfrak{E}$ be the subset of $\mathfrak{D}$ defined by $\varphi(t, \boldsymbol{x}) \leq 0$. If for every point in $\mathfrak{E}$ where $\varphi(t, \boldsymbol{x}) = 0$, the inequality

$$\overline{\boldsymbol{D}}_{+[f]}\varphi(t, \boldsymbol{x}) < 0$$

holds, then $\mathfrak{E}$ is a right majorant for (1) in $\mathfrak{D}$.

As a special case, we have:

**Theorem 4**. Let $\boldsymbol{f}(t, \boldsymbol{x})$ be continuous in $a \leq t < b, |\boldsymbol{x}| < +\infty$ and satisfy the inequality

$$S(\boldsymbol{f}(t, \boldsymbol{x})) < \underline{\boldsymbol{D}}_{+}\omega(t)$$

for $a \leq t < b$, where $S(\boldsymbol{x}) = \omega(t)$ and $\omega(t)$ is a continuous function in $[a, b)$ and $S(\boldsymbol{x})$ is a function of class $(L)$ such that for any right differentiable $\boldsymbol{x}(t)$,

$$D_{+}S\big(\boldsymbol{x}(t)\big) \leq S\big(D_{+}\boldsymbol{x}(t)\big)$$

holds [6], for example, $S(\boldsymbol{x}) = |\boldsymbol{x}|$ or $S(\boldsymbol{x}) = \max\{x_1, \ldots, x_k\}$, etc., then the region $\mathfrak{E}$ defined by $S(\boldsymbol{x}) \leq \omega(t)$ is a majorant to the right for (1).

**Proof**. It is sufficient to set

$$\varphi(t, \boldsymbol{x}) = S(\boldsymbol{x}) - \omega(t).$$

## IV. Conditions of the Majorant Set Using Lower Integrals

**Definition 3**. A real-valued function $\varphi(t, \boldsymbol{x})$ on $\mathfrak{D}$ is called a lower integral of (1) if $\varphi$ belongs to class $(L)$ and for every solution $\boldsymbol{x}(t)$ of (1), $\varphi(t, \boldsymbol{x}(t))$ decreases monotonically in the extended sense[3].

$\varphi(t, \boldsymbol{x})$ is a lower integral of (1) if, and only if, $\varphi$ belongs to class $(L)$ and satisfies the inequality

$$\overline{\boldsymbol{D}}_{+[f]}\varphi(t, \boldsymbol{x}) \leq 0.$$

**Lemma 1**. Let $\varphi(t, \boldsymbol{x}, P)$ be a continuous function of $(t, \boldsymbol{x}, P)$ in a domain $(t, \boldsymbol{x}) \in \mathfrak{D}, P \in \mathfrak{M}$, where $\mathfrak{M}$ is a compact set[4]. If $\varphi(t, \boldsymbol{x}, P)$ is a lower integral of (1) in $\mathfrak{D}$ and belongs to class $(L, 1)$, then $\max_{P\in\mathfrak{M}} \varphi(t, \boldsymbol{x}, P)$ and $\min_{P\in\mathfrak{M}} \varphi(t, \boldsymbol{x}, P)$ are lower integrals of (1) in $\mathfrak{D}$.

**Proof**: Left to the reader.

**Theorem 5**. Let $\mathfrak{D}$ be closed in an open set $\mathfrak{D}^*$ (in the $(t, \boldsymbol{x})$-space) and admissible on both sides for (1), where $\boldsymbol{f}(t, \boldsymbol{x})$ is continuous on $\mathfrak{D}$. A closed set $\mathfrak{E}$ in $\mathfrak{D}$ is a right majorant in $\mathfrak{D}$ if, and only if, there exists a lower

---

[2] δ means an independent positive number from $x(t)$

[3] i.e., $t_1 < t_2$ implies $\phi\big(t_1, x(t_1)\big) \geq \phi\big(t_2, x(t_2)\big)$

[4] i.e., every infinite subset of $\mathfrak{M}$ has at least one limit point in $\mathfrak{M}$

integral $\varphi(t, \boldsymbol{x})$ of (1) in a neighborhood of every point in $\mathfrak{E}$ such that $\varphi(t, \boldsymbol{x}) = 0$ for $(t, \boldsymbol{x}) \in \mathfrak{E}$ and $\varphi(t, \boldsymbol{x}) > 0$ for $(t, \boldsymbol{x}) \in \mathfrak{D} - \mathfrak{E}$.

**Proof**: Since the sufficiency of the conditions is easy to prove, we will only prove the necessity of these conditions.

Let $(a, \boldsymbol{b})$ be a point in $\mathfrak{E}$. Then, there exist positive numbers $l$ and $M$ such that the region $|t - a| \leq l, |\boldsymbol{x} - \boldsymbol{b}| \leq Ml$ is entirely contained in $\mathfrak{D}^*$ and $|\boldsymbol{f}(t, \boldsymbol{x})| \leq M$ holds there. Let $\mathfrak{D}_1$ be the intersection of $\mathfrak{D}$ with this region. For any two points $P = (t_P, \boldsymbol{x_P})$ and $Q = (t_Q, \boldsymbol{x_Q})$ such that $t_P \leq t_Q$, we define the Okamura function $D(P, Q)$ as follows: We divide the interval $[t_P, t_Q]$ into $t_1, t_2, \dots, t_{n-1}$ such that $t_{i-1} \leq t_i$, $t_0 = t_P$, and $t_n = t_Q$. Let $P_i = (t_i, \boldsymbol{x}_i)$ and $Q_i = (t_i, \boldsymbol{x}'_i)$ be points in $\mathfrak{D}_1$ on the same hyperplane $t = t_i$ such that $Q_{i-1}$ and $P_i$ lie on an integral curve running through $\mathfrak{D}_1$, with $P_0 = P$ and $Q_n = Q$, see Fig. 1. $D(P, Q)$ is defined as the infimum of the values $\sum_{i=0}^{n} |\boldsymbol{x}_i - \boldsymbol{x}'_i|$ for all possible points $P_i$ and $Q_i$, where $n$ can vary arbitrarily.

As can be easily proven, the Okamura function $D(P, Q)$ satisfies the following relations:

i) $D(P, Q) \geq 0$.

ii) $D(P, R) \leq D(P, Q) + D(Q, R)$ if $t_P \leq t_Q \leq t_R$.

iii) $|D(P, Q) - D(P', Q')| \leq |\boldsymbol{x}_P - \boldsymbol{x}_{P'}| + |\boldsymbol{x}_Q - \boldsymbol{x}_{Q'}| + M(|t_P - t_{P'}| + |t_Q - t_{Q'}|)$.

iv) If $t_P = t_Q$, then $D(P, Q) = |\boldsymbol{x}_P - \boldsymbol{x}_Q|$.

v) $D(P, Q) = 0$ if, and only if, $P$ and $Q$ lie on an integral curve running through $\mathfrak{D}_1$.

vi) $D(P, X)$ is a lower integral of (1) as a function of $X = (t, \boldsymbol{x})$ for $t \geq t_P$.

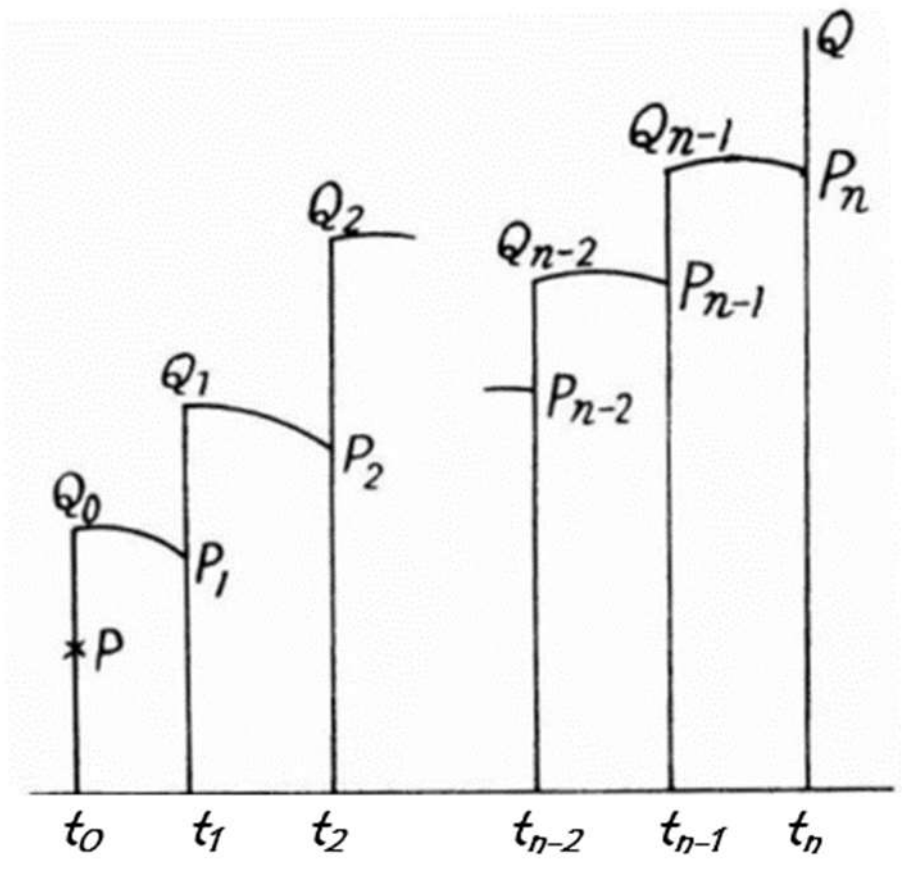

Fig. 1

For $t < t_P$, we extend $D(P, X)$ by

$$D^*(P, X) = \begin{cases} D(P, X) \text{ for } t \geq t_P \\ |\boldsymbol{x} - \boldsymbol{x}_P| + M(t_P - t) \text{ for } t < t_P. \end{cases}$$

$D^*(P, X)$ is then a continuous function of $(P, X)$ for $P \in \mathfrak{D}_1, X \in \mathfrak{D}_1$, belongs to class $(L, 1)$ as a function of $X$, and is a lower integral of (1).

Now we define $\varphi(X)$ as

$$\varphi(X) = \min_{P \in \mathfrak{E}} D^*(P, X),$$

where $\mathfrak{E}_1$ is the intersection of $\mathfrak{E}$ with $\mathfrak{D}_1$. According to Lemma 1, $\varphi(X)$ is also a lower integral of (1). Since $\mathfrak{E}$ is a right majorant, $\varphi(X) = 0$ if, and only if, $X \in \mathfrak{E}_1$, and $\varphi(X) > 0$ for $X \in \mathfrak{D}_1 - \mathfrak{E}_1$. QED.

As an application, we obtain:

**Theorem 6**. Let $\mathfrak{E}$ be a closed set in an open domain $\mathfrak{D}$ and admissible to the right for (1). Suppose the inequality

$$|\boldsymbol{f}(t,\boldsymbol{x}) - \boldsymbol{f}(t,\boldsymbol{x}^*)| \le K|\boldsymbol{x} - \boldsymbol{x}^*| \tag{4}$$

holds for every point $(t,\boldsymbol{x})$ in $\mathfrak{D} - \mathfrak{E}$ and the point $(t,\boldsymbol{x}^*)$ in $\mathfrak{E}$ with the same value of $t$, such that $|\boldsymbol{x} - \boldsymbol{x}^*|$ is minimized. Then $\mathfrak{E}$ in $\mathfrak{D}$ is a right majorant for (1).

**Proof**. Let $(a,\boldsymbol{b})$ be an arbitrary point in $\mathfrak{E}$. Then there exist positive numbers $l$ and $M$ such that the region $\mathfrak{D}_1$: $|t-a| \le l, |\boldsymbol{x} - \boldsymbol{b}| \le Ml$ is contained in $\mathfrak{D}$ and $|\boldsymbol{f}(t,\boldsymbol{x})| < M$ holds. For a fixed $(t,\boldsymbol{x}) \in \mathfrak{D}_1 - \mathfrak{E}$ and a varying $(t^*,\boldsymbol{x}^*) \in \mathfrak{E} \cap \mathfrak{D}_1$, the expression $|\boldsymbol{x} - \boldsymbol{x}^*| + M|t - t^*|$ is minimized only for $t \le t^*$. Therefore, we define

$$\psi(t,\boldsymbol{x}) = \min_{(t^*,\boldsymbol{x}^*)\in\mathfrak{E}_1} [|\boldsymbol{x} - \boldsymbol{x}^*| + M|t - t^*|], \tag{5}$$

where $\mathfrak{E}_1 = \mathfrak{E} \cap \mathfrak{D}_1$. It is clear that $\psi(t,\boldsymbol{x}) = 0$ for $(t,\boldsymbol{x}) \in \mathfrak{E}_1$ and $\psi(t,\boldsymbol{x}) > 0$ for $(t,\boldsymbol{x}) \in \mathfrak{D}_1 - \mathfrak{E}_1$.

If $|\boldsymbol{x} - \boldsymbol{x}^*| + M|t - t^*|$ is minimized for $t^* > t$, $(t^*,\boldsymbol{x}^*) \in \mathfrak{E}_1$ at a fixed $(t,\boldsymbol{x})$, then for $0 < h < t^* - t$,

$$|\boldsymbol{x} + h\boldsymbol{f}(t,\boldsymbol{x}) - \boldsymbol{x}^*| + M\,|t^* - (t+h)| < |\boldsymbol{x} - \boldsymbol{x}^*| + M\,|t - t^*|,$$

i.e., $\psi(t+h, \boldsymbol{x} + h\boldsymbol{f}) < \psi(t,\boldsymbol{x})$. Therefore,

$$\overline{D}_{+[f]}\psi(t,\boldsymbol{x}) \le 0.$$

On the other hand, if $|\boldsymbol{x} - \boldsymbol{x}^*| + M|t^* - t|$ is minimized for $t = t^*$, $(t^*,\boldsymbol{x}^*) \in \mathfrak{E}_1$ at a fixed $(t,\boldsymbol{x})$, then for sufficiently small $h > 0$,

$$|\boldsymbol{x} + h\boldsymbol{f}(t,\boldsymbol{x}) - \boldsymbol{x}^*(t+h)| \le |\boldsymbol{x} - \boldsymbol{x}^*| + h\,[|\boldsymbol{f}(t,\boldsymbol{x}) - \boldsymbol{f}(t,\boldsymbol{x}^*)| + \delta(h)],$$

where $\boldsymbol{x} = \boldsymbol{x}^*(t)$ is an integral curve passing through $(t^*,\boldsymbol{x}^*)$ and lying in $\mathfrak{E}$ for $t^* \le t < t^* + \varepsilon$, and $\delta(h)$ approaches zero as $h$ approaches zero. Therefore,

$$\psi(t+h, \boldsymbol{x} + h\boldsymbol{f}) - \psi(t,\boldsymbol{x}) \le h\,[|\boldsymbol{f}(t,\boldsymbol{x}) - \boldsymbol{f}(t,\boldsymbol{x}^*)| + \delta(h)]$$

which implies, according to (4) and (5), where $t = t^*$,

$$\overline{D}_{+[f]}\psi(t,\boldsymbol{x}) \le K\psi(t,\boldsymbol{x}).$$

Now we set $\varphi(t,\boldsymbol{x}) = \psi(t,\boldsymbol{x})e^{-Kt}$, so $\varphi(t,\boldsymbol{x})$ belongs to class $(L)$ and satisfies the conditions in $\mathfrak{D}_1$: $\overline{D}_{+[f]}\varphi(t,\boldsymbol{x}) \le 0, \varphi(t,\boldsymbol{x}) = 0$ for $(t,\boldsymbol{x}) \in \mathfrak{E}_1$, and $\varphi(t,\boldsymbol{x}) > 0$ for $(t,\boldsymbol{x}) \in \mathfrak{D}_1 - \mathfrak{E}_1$. QED.

## V. Comparison of a System of Equations with a Single Equation

**Lemma 2**. Let $F(t,x)$ be continuous in the region $\mathfrak{D}$: $a \le t < b, x < \Lambda(t)$, and let the region $\mathfrak{E}$: $a \le t < b, x \le \omega(t)$ be a right majorant for

$$\frac{dx}{dt} = F(t,x), \tag{6}$$

where $\Lambda(t)$ and $\omega(t)$ are continuous in $a \le t < b$ and $\omega(t) < \Lambda(t)$. Then there exists a lower integral $\varphi(t,x)$ of (6) in $\mathfrak{D}_1$: $a \le t \le b_1 < b, x \le \Lambda_1(t)$, where $\Lambda_1(t)$ is continuous in $[a,b)$ and $\omega(t) < \Lambda_1(t) < \Lambda(t)$, such that $\varphi(t,x) > 0$ for $x > \omega(t)$, $\varphi(t,x) = 0$ for $x \le \omega(t)$, and $\varphi(t,x)$ increases monotonically with $x$ in the extended sense.

**Proof**. There exists a constant $M > 0$ such that $|F(t,x)| < M$ in $\mathfrak{D}_1$. Let $D(P,X)$ be the Okamura function of (6) in $\mathfrak{D}_1$, where $P = (\xi,\eta)$, $X = (t,x)$, and $\xi \le t$. If $\eta \le \omega(\xi)$ and $x \ge \omega(t)$, then $D(P,X)$ increases monotonically with $x$ in the extended sense.

To see this, let $X = (t,x)$ and $\tilde{X} = (t,\tilde{x})$ be points with the same value of $t$, such that $t > \xi$ and $\tilde{x} > x \ge \omega(t)$. For any $\varepsilon > 0$, there exists a sequence of points $\{P_i\}$ and $\{Q_i\}$ as in the proof of Theorem 5, such that $P_0 = P$, $Q_n = \tilde{X}$, and

$$\sum_{i=0}^{n} \overline{P_i Q_i} < D(P,\tilde{X}) + \varepsilon.$$

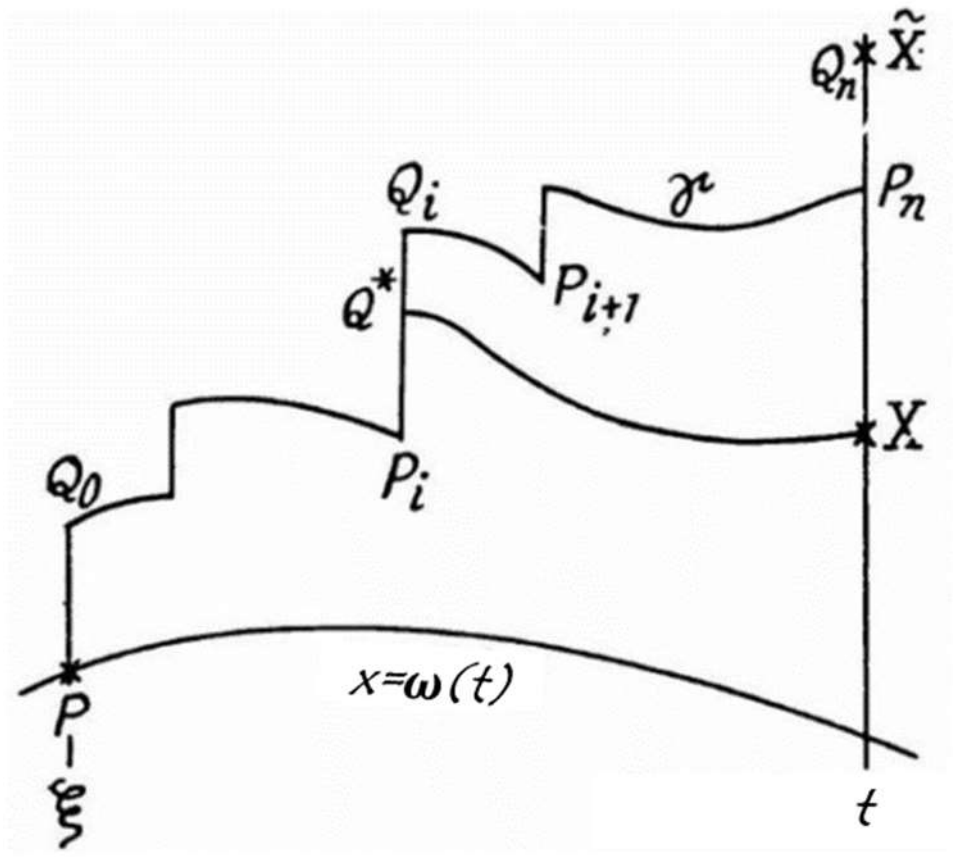

Fig. 2

Let $\mathfrak{S}$ be the curve consisting of segments $Q_iP_{i+1}$ of the integral curves and vertical segments $P_iQ_i$. An integral curve through $X$ intersects $\mathfrak{S}$ at a point $Q^*$. $Q^*$ is either on a segment $P_iQ_i$ or on a segment $Q_iP_{i+1}$ of the integral curve. Then we have

$$D(P,X) \leq \overline{P_0Q_0} + \cdots + \overline{P_iQ_i} < D(P,\tilde{X}) + \varepsilon.$$

Hence, $D(P,X) \leq D(P,\tilde{X})$.

For $P \in \mathfrak{E}_1$ and $X \in \mathfrak{D}_1 - \mathfrak{E}_1$, we set

$$D^*(P,X) = \begin{cases} D(P,X) \text{ if } t \geq \xi \\ x - \omega(t) + 2M(\xi - t) \text{ if } t < \xi. \end{cases}$$

We then define $\varphi(X)$ $as$

$$\varphi(X) = \begin{cases} \min\limits_{P\in\mathfrak{E}_1} D^*(P,X) \ \text{ if } X \in \mathfrak{D}_1 - \mathfrak{E}_1 \\ 0 \text{ if } X \in \mathfrak{E}_1. \end{cases}$$

Thus, $\varphi(X)$ possesses all the properties mentioned in the theorem. QED.

**Theorem 7**. Let $F(t,x)$ be continuous in the range $a \leq t < b$, $-\infty < x < \infty$, and let the range $a \leq t < b$, $x \leq \omega(t)$ be a right majorant for the equation

$$\frac{dx}{dt} = F(t,x), \tag{6}$$

where $\omega(t)$ is continuous in the range $a \leq t < b$.

Now let $\boldsymbol{f}(t,\boldsymbol{x})$ be continuous in the range $a \leq t < b$, $|\boldsymbol{x}| < +\infty$, and let $S(t,\boldsymbol{x})$ be of class $(L)$ with the property:

$$\overline{D}_{+[\boldsymbol{f}]}S(t,\boldsymbol{x}) \leq F(t,S(t,\boldsymbol{x})). \tag{7}$$

Then, $\mathfrak{E}$ defined by $a \leq t < b$, $S(t,\boldsymbol{x}) \leq \omega(t)$ is a right majorant for the equation

$$\frac{d\boldsymbol{x}}{dt} = \boldsymbol{f}(t,\boldsymbol{x}). \tag{1}$$

**Proof**. According to Lemma 2, there exists a function $\varphi(t,x)$ of class $(L)$ in the range $a \leq t \leq b_1 < b$, $|x| \leq \Lambda$, such that $\varphi(t,x) < 0$ for $x > \omega(t)$ and $\varphi(t,x) = 0$ for $x \leq \omega(t)$,

$$\overline{D}_{+[F]}\varphi(t,x) \leq 0 \tag{8}$$

and $\varphi(t,x)$ monotonically increases with $x$ in the extended sense. Let $\Phi(t,\boldsymbol{x}) = \varphi(t,S(t,\boldsymbol{x}))$. Then, $\Phi(t,\boldsymbol{x})$ belongs to class $(L)$. According to (7), for $h > 0$,

$$S(t+h, \boldsymbol{x}+h\boldsymbol{f}) \leq S(t, \boldsymbol{x}) + h[F(t, S(t, \boldsymbol{x}) + \delta(h)],$$

where $\delta(h)$ approaches zero as $h$ approaches zero. Therefore,

$$\Phi(t+h, \boldsymbol{x}+h\boldsymbol{f}) - \Phi(t, \boldsymbol{x}) \leq \varphi(t+h, S+hF) - \varphi(t, S) + \alpha h \delta(h),$$

thus, according to (8),

$$\overline{D}_{+[f]}\Phi(t, \boldsymbol{x}) \leq 0.$$

Thus, $\Phi(t, \boldsymbol{x})$ is a lower integral of (1), $\Phi(t, \boldsymbol{x}) = 0$ for $(t, \boldsymbol{x}) \in \mathfrak{E}$, and $\Phi(t, \boldsymbol{x}) > 0$ for $(t, \boldsymbol{x}) \in \mathfrak{D} - \mathfrak{E}$. $\mathfrak{E}$ is then a right majorant. QED.

It follows as special case of 7:

**Theorem 8**. Let $F(t, x)$ and $\boldsymbol{f}(t, \boldsymbol{x})$ be functions with the same properties as in Theorem 7, while the inequality (7) is replaced by

$$S(\boldsymbol{f}(t, \boldsymbol{x})) \leq F(t, S(\boldsymbol{x})), \tag{9}$$

where $S(\boldsymbol{x})$ belongs to class $(L)$ and satisfies

$$D_+ S(\boldsymbol{x}(t)) \leq S\,(D_+ \boldsymbol{x}(t))$$

for any right-differentiable function $\boldsymbol{x}(t)$, for example, $\boldsymbol{S}(\boldsymbol{x}) = |\boldsymbol{x}|$.

If the range $a \leq t < b, x \leq \omega(t)$ is a right majorant for (6), where $\omega(t)$ is continuous in the range $a \leq t < b$, then the range defined by $a \leq t < b$, $S(\boldsymbol{x}) \leq \omega(t)$ is a right majorant for (1).

**Mathematical Institute of the Imperial University of Osaka.**

## Acknowledgment

An enhanced pdf format of the Nagumo's paper in German and its initial English translation were produced by the Boeing Company, Germany. Another English version of the paper was independently created by Prof. Dr.-Ing. Florian Holzapfel and his PhD student Tugba Akman. Many thanks go to all who have helped bringing Nagumo's seminal work to a broader audience.

## APPENDIX A: GERMAN VERSION OF ORIGINAL PAPER

In the following, a German version with improved pdf format readability of Nagumo's paper is attached as a reference. Note that this version does not include the notation changes above.

# Über die Lage der Integralkurven gewöhnlicher Differentialgleichungen

## § 1. Einleitung.

In dieser Note werden $k$-dimensionale Vektoren mit dicken Buchstaben bezeichnet. Wir sollen also unter

$$\frac{d\boldsymbol{y}}{dx} = \boldsymbol{f}(x, \boldsymbol{y}) \tag{1}$$

ein System der Differentialgleichungen

$$\frac{dy_i}{dx} = f_i\left(x, y_1, \cdots, y_k\right)$$
$$(i = 1, 2, \cdots, k)$$

verstehen.

*O. Perron* hat den Existenzbeweis der Lösungen einer gewöhnlicher Differentialgleichung $\frac{dy}{dx} = F(x, y)$ in der Form gegeben, dass sie in einen Bereich $a \leqq x < b, \omega_1(x) \leqq y \leqq \omega_2(x)$ eingeschlossen werden, wobei $\omega_i(x)$ den Bedingungen

$$D_{\pm}\omega_1(x) \leqq F\left(x, \omega_1(x)\right), \quad D_{\pm}\omega_2(x) \geqq F\left(x, \omega_2(x)\right)$$

genügen[1]. Diese Hinsicht ist von *Hukuhara* äusserst ausgeführt[2]. Er nennt auch eine Teilmenge $\mathfrak{E}$ der Punktmenge $\mathfrak{D}$ des $(x, \boldsymbol{y})$-Raumes "*nach rechts majorant in* $\mathfrak{D}$,"wenn jede in $\mathfrak{D}$ liegende Lösungskurve von (1) mit einem beliebigen Anfangspunkt $(x_0, \boldsymbol{y}_0)$ in $\mathfrak{E}$ immer in $\mathfrak{E}$ bleibt für $x \geqq x_0$.

Das Hauptziel der vorliegenden Note ist in einem Sinne notwendige und hinreichende Bedingungen zu geben, dass $\mathfrak{E}$ in $\mathfrak{D}$ nach rechts majorant ist. Die Bedingungen werden mittels Unterintegrals[3] gegeben, das sich auf der Idee von *Okamura* beruht, die er für die Unitätsbedingung der Lösung von (1) gebraucht hat[4].

[1] Math. Ann. 76 (1915).

[2] Nippon Sugaku-Buturigakkwai Kwaisi 5 (1931) u. 6(1932) (japanisch). Vgl. Memoirs of the Fac. of Sci. Kyūsyū Imp. Univ. Ser. A. 2. (1941) 1.-25.

[3] Die Definition wird in $4 dieser Note gegeben.

[4] Memoirs of the College of Sci. Kyoto Imp. Univ. Ser. A. 23 (1941) 225-231.

## § 2. Zülassige Menge.

**Definition 1.** *Eine Punktmenge* $\mathfrak{M}$ *des* $(x, \boldsymbol{y})$*-Raumes heisst* nach rechts zulässig *für die Differentialgleichung (1), wenn es für jeden Punkt* $(x_0, \boldsymbol{y}_0)$ *von* $\mathfrak{M}$ *eine positive Zahl* $l$ *gibt, sodass eine in* $\mathfrak{M}$ *liegende Integralkurve von (1) mit dem Anfangspunkt* $(x_0, \boldsymbol{y}_0)$ *existiert mindestens für* $x_0 \leqq x < x_0 + l$.

Linksseitige Zulässigkeit kann man ganz analog definieren.

**Satz 1.** *Es sei* $\mathfrak{D}$ *eine offene Menge im Raum von* $(x, \boldsymbol{y})$ *und sei* $\mathfrak{E}$ *eine in* $\mathfrak{D}$ *abgeschlossene Menge, worauf* $\boldsymbol{f}(x, \boldsymbol{y})$ *stetig ist.* $\mathfrak{E}$ *ist dann und nur dann nach rechts zulässig für (1), wenn es für jeden Punkt* $(x_0, \boldsymbol{y}_0)$ *von* $\mathfrak{E}$ *und eine beliebige positive Zahl* $\varepsilon$ *einen Punkt* $(x_1, \boldsymbol{y}_1)$ *von* $\mathfrak{E}$ *gibt mit der Beschaffenheit:* $x_0 < x_1 < x_0 + \varepsilon$ *und*

$$\left| \frac{\boldsymbol{y}_1 - \boldsymbol{y}_0}{x_1 - x_0} - \boldsymbol{f}(x_0, \boldsymbol{y}_0) \right| < \varepsilon$$

*Beweis.* Es braucht nur die Hinlänglichkeit der Bedingungen zu beweisen, weil die Notwendigkeit klar ist. Für einen Punkt $(x_0, \boldsymbol{y}_0)$ von $\mathfrak{E}$ gibt es positive Zahlen $l$ und $M$ derart, dass der Bereich $\mathfrak{D}_1 : x_0 \leqq x \leqq x_0 + l, |\boldsymbol{y} - \boldsymbol{y}_0| \leqq (M+1)l$ in $\mathfrak{D}$ liegt, und in $\mathfrak{D}_1$ $|\boldsymbol{f}(x, \boldsymbol{y})| \leqq M$ ist. $P_0, P_1, P_2, \ldots, P_n = (x_n, \boldsymbol{y}_n)$ sei eine Punktfolge mit den Bedingungen: $x_{\nu-1} < x_\nu < x_{\nu-1} + \varepsilon, P_\nu \in \mathfrak{E}$ und

$$\left| \frac{\boldsymbol{y}_\nu - \boldsymbol{y}_{\nu-1}}{x_\nu - x_{\nu-1}} - \boldsymbol{f}\left(x_{\nu-1}, \boldsymbol{y}_{\nu-1}\right) \right| < \varepsilon \quad (\varepsilon < 1) \tag{2}$$

Es sei $\mathfrak{M}$ die Menge aller möglichen solchen Punkte $P_n$. Die obere Grenze $\xi$ der Werte von $x$, für die $(x, \boldsymbol{y}) \in \mathfrak{M}$ sind, genügt $\xi > x_0 + l$. Denn, wäre dies nicht der Fall, würde $\mathfrak{M}$ in $\mathfrak{D}_1$ enthalten. Es gibt einen Häufungspunkt $(\xi, \boldsymbol{y}^*)$ von $\mathfrak{M}$ auf $x = \xi$. Da $(\xi, \boldsymbol{y}^*) \in \mathfrak{E}$ ist, so gibt es einen $(\xi_1, \boldsymbol{y}_1^*) \in \mathfrak{E}$ derart, dass $\xi < \xi_1 < \xi + \varepsilon$ und

$$\left| \frac{\boldsymbol{y}_1^* - \boldsymbol{y}^*}{\xi_1 - \xi} - \boldsymbol{f}(\xi, \boldsymbol{y}^*) \right| < \varepsilon. \tag{3}$$

Es gibt aber eine endliche Folge $P_0, P_1, \ldots, P_n$ mit den Bedingungen $x_{\nu-1} < x_\nu < x_{\nu-1} + \varepsilon$ und (2) aus $\mathfrak{M}$, sodass $P_n$ in die beliebige Nähe von $(\xi, \boldsymbol{y}^*)$ kommt. Also $x_n < \xi_1 < x_n + \varepsilon$ und nach (3)

$$\left| \frac{\boldsymbol{y}_1^* - \boldsymbol{y}_n}{\xi_1 - x_n} - \boldsymbol{f}(x_n, \boldsymbol{\eta}_n) \right| < \varepsilon$$

Folglich $P_{n+1} = (\xi_1, \boldsymbol{y}_1^*) \in \mathfrak{M}$ mit $\xi_1 = x_{n+1} > \xi$, gegen der Definition von $\xi$.

Nun sei $\boldsymbol{y} = \boldsymbol{Y}_N(x)$ die Gleichung des Streckenzuges, der eine Punktfolge $P_0, P_1, \ldots, P_n$ mit den Bedingungen $P_\nu \in \mathfrak{E}$, $x_{\nu-1} < x_\nu < x_{\nu-1} + \varepsilon$ und (2) verbindet, wobei $\varepsilon = \frac{1}{N}$ ist. Die Kurven $\boldsymbol{y} = \boldsymbol{Y}_N(x)$ liegen für $x_0 \leqq x \leqq x_0 + l$ immer in $\mathfrak{D}_1$ und genügen der Ungleichung

$$|\boldsymbol{Y}_N(x') - \boldsymbol{Y}_N(x'')| \leqq (M+1)\,|x' - x''|\,.$$

Es gibt dann eine Teilfolge $\{N_i\}$ der natürlichen Zahlen, sodass $\boldsymbol{Y}_{N_i}(x)$ für $N_i \to \infty$ in $\langle x_0, x_0 + l\rangle$ gleichmässig gegen eine stetige Kurve $\boldsymbol{y} = \boldsymbol{Y}(x)$ konvergiert, die in $\mathfrak{E}$ liegt. Man kann nicht schwer beweisen, dass für genügend grosse $N_i$

$$\left|\frac{\boldsymbol{Y}_{N_i}(x') - \boldsymbol{Y}_{N_i}(x)}{x' - x} - \boldsymbol{f}(x, \boldsymbol{Y}(x))\right| < \varepsilon$$

ist, wenn $|x' - x| < \delta, x_0 \leqq x \leqq x_0 + l$, wobei $\delta > 0$ genügend klein ist. Daraus fo'gt für $N_i \to \infty$ und dann für $\delta \to 0$, dass

$$\frac{d}{dx}\boldsymbol{Y}(x) = \boldsymbol{f}(x, \boldsymbol{Y}(x))$$

für $x_0 \leqq x < x_0 + l$ und $\boldsymbol{Y}(x_0) = \boldsymbol{y}_0$. W.z.b.w. □

**Satz 2.** *Es seien $\mathfrak{D}$ und $\mathfrak{E}$ von denselben Bedeutungen wie in Satz 1, Ist $\mathfrak{E}$ nach rechts zulässig für (1), so kann jede Integralkurve von (1), die in $\mathfrak{E}$ liegt, bis auf den Rand von $\mathfrak{D}$ fortsetzbar.*

*Beweis.* Dem Leser überlassen. □

Als eine Anwendung von Satz 1 und Satz 2 bekommen wir nicht schwer:

*Es sei $f(x,y)$ im Bereiche $a \leqq x < b$, $\omega_1(x) \leqq y \leqq \omega_2(x)$ stetig, wobei $\omega_i(x)$ in $a \leqq x < b$ stetig sind und genügen den Relationen*

$$\underline{\boldsymbol{D}}_+\omega_1(x) \leqq f(x, \omega_1(x))\,, \quad \overline{\boldsymbol{D}}_+\omega_2(x) \geqq f(x, \omega_2(x))\,.$$

*Es gibt dann mindestens eine in $a \leqq x < b$ stetige Lösung $y = y(x)$ von $\frac{dy}{dx} = f(x,y)$ mit den Bedingungen $y(a) = y_0$, $(\omega_1(a) \leqq y_0 \leqq \omega_2(a))$, und*

$$\omega_1(x) \leqq y(x) \leqq \omega_2(x)$$

*für $a \leqq x < b$.*

## §3. Operation $\overline{\boldsymbol{D}}_{+[f]}\varphi$.

**Definition 2.** *Eine auf einer Menge $\mathfrak{D}$ im $(x, \boldsymbol{y})$-Raum definierte Funktion $\varphi(x, \boldsymbol{y})$ heisst von der Klasse $(L)$, genauer von der Klasse $(L, \alpha)$, in $\mathfrak{D}$, wenn sie in $\mathfrak{D}$ stetig ist und es eine Konstante $\alpha$ gibt, sodass für beliebige $(x, \boldsymbol{y}_i) \in \mathfrak{D}$ $(i = 1, 2)$ mit einem gemeinsamen Wert von $x$ die Ungleichung besteht:*

$$|\varphi(x, \boldsymbol{y}_1) - \boldsymbol{\varphi}(x, \boldsymbol{y}_2)| \leqq \boldsymbol{\alpha}\,|\boldsymbol{y}_1 - \boldsymbol{y}_2|\,.$$

Ist $\varphi(x, \boldsymbol{y})$ eine reellwertige Funktion der Klasse $(L)$ auf $\mathfrak{D}$, so hat der Grenzwert

$$\overline{\lim_{h\to+0}} \frac{\varphi\left(x_0+h, \boldsymbol{y}\left(x_0+h\right)\right)-\varphi\left(x_0, \boldsymbol{y}_0\right)}{h}$$

immer denselben Wert, wenn nur $\boldsymbol{y}(x)$ eine beliebige Funktion derart ist, dass $(x, \boldsymbol{y}(x)) \in \mathfrak{D}$ für $x_0 < x < x_0 + \delta^5$ und

$$\lim_{h\to+0} \frac{\boldsymbol{y}\left(x_0+h\right)-\boldsymbol{y}_0}{h} = \boldsymbol{f}\left(x, \boldsymbol{y}_0\right)$$

ist. *Diesen Grenzwert bezeichnen wir dann mit* $\boldsymbol{D}_{+[f]}\varphi\left(x_0, \boldsymbol{y_0}\right)$. Also

$$\overline{\boldsymbol{D}}_{+[f]}\varphi\left(x_0, \boldsymbol{y}_0\right) = \overline{\lim_{h\to+0}} \frac{\varphi\left(x_n+h, \boldsymbol{y_0}+h\boldsymbol{f}_0\right)-\varphi\left(x_0, \boldsymbol{y_0}\right)}{h},$$

wenn nur $\left(x_0+h, \boldsymbol{y}_0+h\boldsymbol{f}_0\right) \in \mathfrak{D}$ für genügend kleine $h \geqq 0$, wobei $\boldsymbol{f}_0 = \boldsymbol{f}\left(x_0, \boldsymbol{y}_0\right)$.

Man kann leicht für die Funktionen der Klasse $(L)$ folgende Relationen beweisen:

$$\begin{aligned}\overline{\boldsymbol{D}}_{+[f]}\left[\varphi_1(x, \boldsymbol{y})+\varphi_2(x, \boldsymbol{y})\right] &\leqq \overline{\boldsymbol{D}}_{+[f]}\varphi_1(x, \boldsymbol{y})+\overline{\boldsymbol{D}}_{+[f]}\varphi_2(x, \boldsymbol{y}). \\ \overline{\boldsymbol{D}}_{+[f]}\left[\varphi_1(x, \boldsymbol{y})\cdot\varphi_2(x, \boldsymbol{y})\right] &\leqq \varphi_1(x, \boldsymbol{y})\cdot\overline{\boldsymbol{D}}_{+[f]}\varphi_2(x, \boldsymbol{y}) \\ &\quad+\varphi_2(x, \boldsymbol{y})\cdot\overline{\boldsymbol{D}}_{+[f]}\varphi_1(x, \boldsymbol{y}),\end{aligned}$$

wenn $\varphi_i(x, \boldsymbol{y}) \geqq 0 (i=1,2)$ sind.

Nicht schwer kann man beweisen folgenden:

**Satz 3.** *Es sei* $\mathfrak{D}$ *eine Menge, worauf* $\boldsymbol{f}(x, \boldsymbol{y})$ *stetig ist, und sei* $\varphi(x, \boldsymbol{y})$ *eine Funktion der Klasse* $(L)$ *auf* $\mathfrak{D}$. $\mathfrak{E}$ *sei die durch* $\varphi(x, \boldsymbol{y}) \leqq 0$ *definierte Teilmenge von* $\mathfrak{D}$. *Besteht für jeden Punkt von* $\mathfrak{E}$, *sodass* $\varphi(x, \boldsymbol{y}) = 0$, *die Ungleichung*

$$\overline{\boldsymbol{D}}_{+[\boldsymbol{f}]}\varphi(x, \boldsymbol{y}) < 0,$$

*so ist* $\mathfrak{E}$ *in* $\mathfrak{D}$ *näch rechts majorant für (1).*

Als einen spezialen Fall erhält mann:

**Satz 4.** *Es sei* $\boldsymbol{f}(x, \boldsymbol{y})$ *in* $a \leqq x < b$, $|\boldsymbol{y}| < +\infty$ *stetig und genüge der Ungleichung*

$$S\left(\boldsymbol{f}(x, \boldsymbol{y})\right) < \underline{\boldsymbol{D}}_{+}\boldsymbol{\omega}(x)$$

*für* $a \leqq x < b$, $S(\boldsymbol{y}) = \boldsymbol{\omega}(x)$, *wobei* $\boldsymbol{\omega}(x)$ *eine in* $\langle a, b)$ *stetige Frunktion und* $S(\boldsymbol{y})$ *eine Funktion der Klasse (L), sodass für eine beliebige nach rechts differentiierbare* $\boldsymbol{y}(x)$

$$D_{+}S(\boldsymbol{y}(x)) \leqq S\left(D_{+}\boldsymbol{y}(x)\right)$$

[5] $\delta$ bedeutet eine von $\boldsymbol{y}(x)$ abhängige positive Zahl.

*ist*[6]*, z.B.* $S(\boldsymbol{y}) = |\boldsymbol{y}|$*, oder* $\boldsymbol{S}(\boldsymbol{y}) = \mathrm{Max}\,[y_1, \ldots ; y_k]$*, udgl. Dann ist der durch* $S(\boldsymbol{y}) \leqq \boldsymbol{\omega}(x)$ *defnierte Bereich* $\mathfrak{E}$ *nach rechts majorant für (1).*

*Beweis.* Man braucht nur zu setzen

$$\varphi(x, \boldsymbol{y}) = S(\boldsymbol{y}) - \omega(x).$$

□

# §4. Bedingungen der Majoranten Menge mittels Unterintegrals.

**Definition 3.** *Eine reellwertige Funktion* $\varphi(x, \boldsymbol{y})$ *auf* $\mathfrak{D}$ *heisst ein Unterintegral von (1), wenn* $\varphi$ *zur Klasse* $(L)$ *gehört und für jede Lösung* $\boldsymbol{y}(x)$ *von (1)* $\varphi(x, \boldsymbol{y}(x))$ *monoton abnimmt im erweiterten Sinne*[7]*.*

$\varphi(x, \boldsymbol{y})$ ist ein Unterintegral von (1) dann und nur dann, wenn $\varphi$ zur Klasse (L) gehört.und genügt der Ungleichung

$$\overline{\boldsymbol{D}}_{+[\boldsymbol{f}]}\varphi(x, \boldsymbol{y}) \leqq 0.$$

**Hilfsatz 1.** *Es sei* $\varphi(x, \boldsymbol{y}, P)$ *eine in einem Bereich* $(x, \boldsymbol{y}) \in \mathfrak{D}$*,* $P \in \mathfrak{M}$ *stetige Funktion von* $(x, \boldsymbol{y}, P)$*, wobei* $\mathfrak{M}$ *eine in sich kompakte Menge ist*[8]*. Ist* $\varphi(x, \boldsymbol{y}, P)$ *ein Unterintegral von (1) in* $\mathfrak{D}$ *und gehört da zur Klasse* $(L, 1)$*, so sind* $\underset{P \in \mathfrak{M}}{\mathrm{Max}}\, \varphi(x, \boldsymbol{y}, P)$ *und* $\underset{P \in \mathfrak{M}}{\mathrm{Min}}\, \varphi(x, \boldsymbol{y}, P)$ *in* $\mathfrak{D}$ *Unterintegrale von (1).*

*Beweis.* Dem Leser überlassen. □

**Satz 5.** *Es sei* $\mathfrak{D}$ *auf einer offenen Menge* $\mathfrak{D}^*$*( im* $(x, \boldsymbol{y})$*-Raum) abgeschlossen und nach beiden Seiten zülassig für (1), wobei* $\boldsymbol{f}(x, \boldsymbol{y})$ *auf* $\mathfrak{D}$ *stetig ist.*

*Eine in* $\mathfrak{D}$ *abgeschlossene Menge* $\mathfrak{E}$ *ist in* $\mathfrak{D}$ *nach rechts majorant dann und nur dann, wenn es in einer Umgebung des jeden Punktes von & ein Unterintegral* $\varphi(x, \boldsymbol{y})$ *von (1) gibt, sodass* $\varphi(x, \boldsymbol{y}) = 0$ *für* $(x, \boldsymbol{y}) \in \mathfrak{E}$ *und* $\varphi(x, \boldsymbol{y}) > 0$ *fur* $(x, \boldsymbol{y}) \in \mathfrak{D} - \mathfrak{E}$*.*

*Beweis.* Da die Hinlänglichkeit der Bedingungen leicht zu beweisen ist, so beweisen wir nur die Notwendigkeit dieser Bedingungen.

Es sei $(a, \boldsymbol{b})$ ein Punkt von $\mathfrak{E}$. Es gibt dann positive Zahlen $l$ und $M$, sodass der Bereich $|x - a| \leqq l, |\boldsymbol{y} - \boldsymbol{b}| \leqq Ml$ ganz in $\mathfrak{D}^*$ liegt und da $|\boldsymbol{f}(x, \boldsymbol{y})| \leqq M$ ist. $\mathfrak{D}_1$ sei der Durchschnitt von $\mathfrak{D}$ mit diesem Bereich. Für beliebige zwei Punkte $P = (x_P, \boldsymbol{y}_P)$ und $Q = \big(x_Q, \boldsymbol{y_Q}\big)$, sodass $x_P \leqq x_Q$, definieren wir die *Okamurasche Funktion* $D(P, Q)$ folgendermassen: Wir teilen das Intervall $\langle x_P, x_Q \rangle$ durch $x_1, x_2, \ldots, x_{n-1}$, sodass $x_{i-1} \leqq x_i, x_0 = x_P$ und $x_n = x_Q$. $P_i = (x_i, \boldsymbol{y}_i)$ und $Q_i = (x_i, \boldsymbol{y}'_i)$ seien Punkte von $\mathfrak{D}_1$ auf derselben Hyperebene $x = x_i$ derart, dass $Q_{i-1}$ und $P_i$ auf einer in $\mathfrak{D}_1$ laufenden Integralkurve liegen, $P_0 = P$ und $Q_n = Q$. (Vgl. Fig. 1.) $D(P,Q)$ sei die untere Genze der Werte $\sum_{i=0}^{n} |\boldsymbol{y_i} - \boldsymbol{y'_i}|$ für alle möglichen solchen Punkte $\boldsymbol{P}_i$ und $Q_i$, wobei $n$ auch beliebig variert.

[6]Vgl. Hukubara: Sur la Fonction $S(x)$ de M.E. Kamke, Jap. Jour. of Math. 17. (1941), 289.
[7]D.h., aus $x_1 < x_2$ folgt, dass $\varphi(x_1, \boldsymbol{y}(x_1)) \geqq \varphi(x_2, \boldsymbol{y}(x_2))$.
[8]D.h., jede unendliche Teilmenge von $\mathfrak{M}$ hat mindestens einen Häufungspunkt in $\mathfrak{M}$.

Wie man nicht schwer beweist, genügt $D(P,Q)$ folgenden Relationen.

i) $D(P,Q) \geqq 0$,

ii) $D(P,R) \leqq D(P,Q) + D(Q,R)$, wenn $x_P \leqq x_Q \leqq x_R$.

iii) $|D(P,Q) - D(P',Q')| \leqq |\boldsymbol{y}_P - \boldsymbol{y}_{P'}| + |\boldsymbol{y}_Q - \boldsymbol{y}_{Q'}|$

$$+M\left(|x_P - x_{P'}| + |x_Q - x_{Q'}|\right).$$

iv) Ist $x_P = x_{\boldsymbol{Q}}$, so ist $D(P,Q) = |\boldsymbol{y}_P - \boldsymbol{y}_{\boldsymbol{Q}}|$.

v) $D(P,Q) = 0$ dann und nur dann, wenn $P$ und $Q$ auf einer in $\mathfrak{D}_1$ laufenden Integralkurve liegen.

vi) $D(P,X)$ ist als eine Funktion von $X = (x,y)$ ein Unterintegral von (1) für $x \geqq x_P$.

Für $x < x_P$ erweitern wir $D(P,X)$ durch

$$D^*(P,X) = \begin{cases} D(P,X) & \text{für} \quad x \geqq x_P, \\ |\boldsymbol{y} - \boldsymbol{y}_P| + M(x_P - x) & \text{für} \quad x < x_P. \end{cases}$$

$D^*(P,X)$ ist dann eine stetige Funktion von $(P,X)$ für $P \in \mathfrak{D}_1, X \in \mathfrak{D}_1$, gehört zur Klasse $(L,1)$ als eine Funkion von $X$ und ist eine Unterintegral von (1).

Nun definieren wir $\varphi(X)$ durch

$$\varphi(X) = \operatorname*{Min}_{P \in \mathfrak{E}_1} D^*(P,X)$$

wobei $\mathfrak{E}_1$ der Durchschnitt von $\mathfrak{E}$ mit $\mathfrak{D}_1$ ist. Nach Hilfassatz 1 ist dann $\varphi(X)$ auch ein Unterintegral von (1). Da $\mathfrak{E}$ nach rechts majorant ist, $\varphi(X) = 0$ dann und nur dann, wenn $X \in \mathfrak{E}_1$, und $\varphi(X) > 0$ für $X \in \mathfrak{D}_1 - \mathfrak{E}_1$. W.z.b.w. □

Als eine Anwendung bekommen wir:

**Satz 6.** *Es sei $\mathfrak{E}$ in einem offenen Gebiet $\mathfrak{D}$ abgeschlossen und nach rechts zulässig für (1). Es bestehe die Ungleichung*

$$|\boldsymbol{f}(x,\boldsymbol{y}) - \boldsymbol{f}(x,\boldsymbol{y}^*)| \leqq K|\boldsymbol{y} - \boldsymbol{y}^*| \tag{4}$$

*für jeden Punkt $(x,\boldsymbol{y})$ von $\mathfrak{D}$-$E$ und den Punkt $(x,\boldsymbol{y}^*)$ von $\mathfrak{E}$ mit demselben Wert von $x$, sodass $|\boldsymbol{y} - \boldsymbol{y}^*|$ minimum ist. Dann ist $\mathfrak{E}$ in $\mathfrak{D}$ nach rechts majorant für (1).*

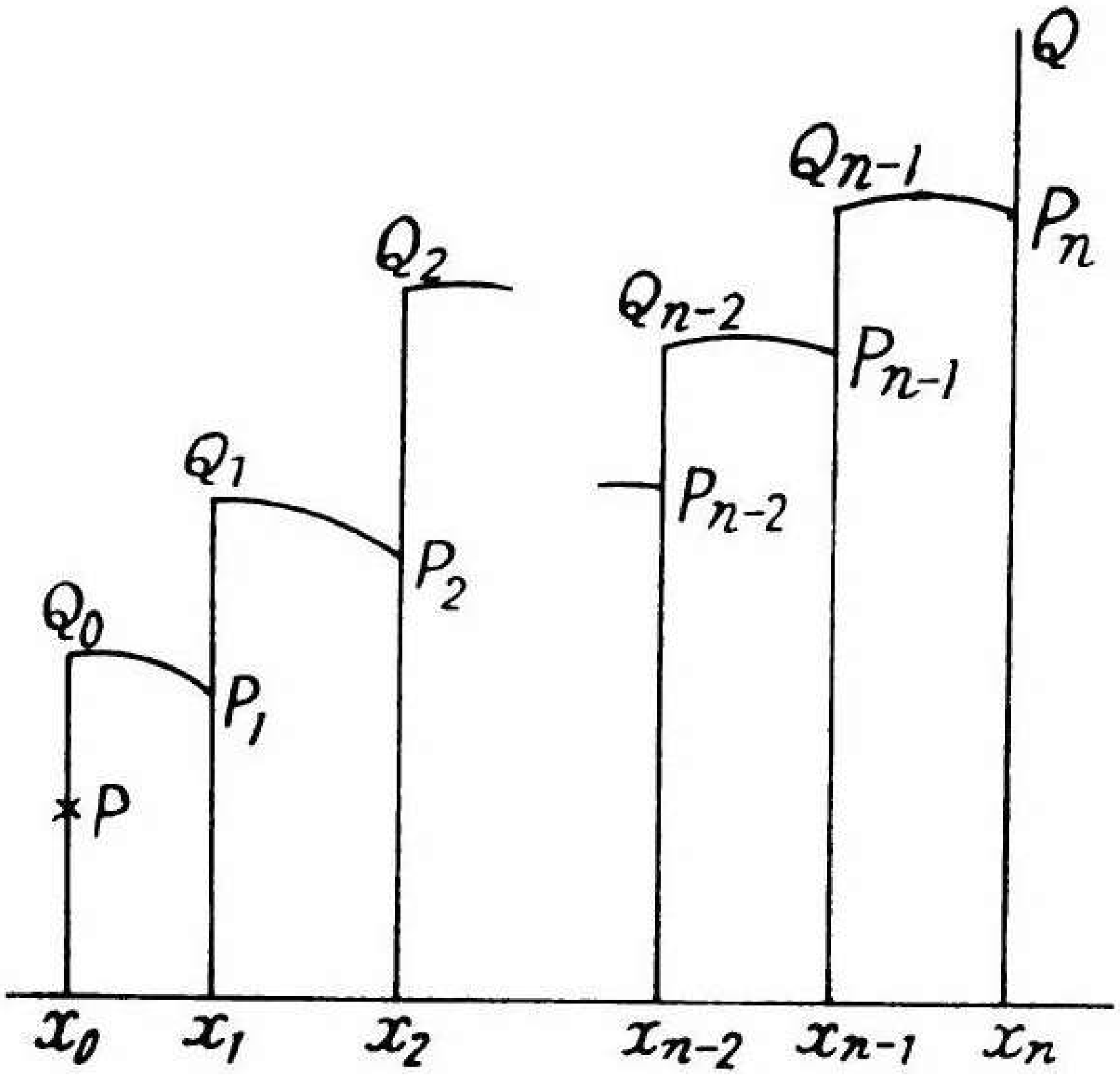

Fig. 1

*Beweis.* Sei $(a, \boldsymbol{b})$ ein beliebiger Punkt von $\mathfrak{E}$, so gibt es positive Zahlen $l$ und $M$, sodass der Bereich $\mathfrak{D}_1 : |x-a| \leqq l, |\boldsymbol{y}-\boldsymbol{b}| \leqq Ml$ in $\mathfrak{D}$ liegt und da $|\boldsymbol{f}(x, \boldsymbol{y})| < M$ ist. Für einen festen $(x, \boldsymbol{y}) \in \mathfrak{D}_1 - \mathfrak{E}$ und einen veränderlichen $(x^*, \boldsymbol{y}^*) \in \mathfrak{E} \cdot \mathfrak{D}_1$ wird $|\boldsymbol{y} - \boldsymbol{y}^*| + M\,|x - x^*|$ minimum nur für $x \leqq x^*$. Wir definieren also

$$\boldsymbol{\psi}(x, \boldsymbol{y}) = \underset{(x^*, \boldsymbol{y}^*) \in \mathfrak{E}_1}{\mathrm{Min}} [|\boldsymbol{y} - \boldsymbol{y}^*| + M\,|x - x^*|] \tag{5}$$

wobei $\mathfrak{E}_1 = \mathfrak{E} \cdot \mathfrak{D}_1$. Es ist klar, dass $\boldsymbol{\psi}(x, \boldsymbol{y}) = 0$ ist für $(x, \boldsymbol{y}) \in \mathfrak{E}_1$ und $\psi(x, \boldsymbol{y}) > 0$ für $(x, \boldsymbol{y}) \in D_1 - \mathfrak{E}_1$.

Wird $|\boldsymbol{y} - \boldsymbol{y}^*| + M\,|x - x^*|$ minimum für $x^* > x$, $(x^*, \boldsymbol{y}^*) \in \mathfrak{E}_1$ bei einem festen $(x, \boldsymbol{y})$, so gilt für $0 < h < x^* - x$,

$|\boldsymbol{y} + h\boldsymbol{f}(x, \boldsymbol{y}) - \boldsymbol{y}^*| + M\,|x^* - (x + h)| < |\boldsymbol{y} - \boldsymbol{y}^*| + M\,|x - x^*|$,

also $\boldsymbol{\psi}\,(x + h, \boldsymbol{y} + h\boldsymbol{f}) < \psi(x, \boldsymbol{y})$, folglich

$$\overline{\boldsymbol{D}}_{+[\boldsymbol{f}]}\boldsymbol{\psi}(x, \boldsymbol{y}) \leqq 0$$

Wird dagegen $|\boldsymbol{y}-\boldsymbol{y}^*| + \boldsymbol{M}\,|x^*-x|$ minimum für $x = x^*$, $(x^*, \boldsymbol{y}^*) \in \mathfrak{E}_1$ beim festen $(x, \boldsymbol{y})$, so gilt für genügend kleine $h > 0$

$$|\boldsymbol{y} + h\boldsymbol{f}(x,\boldsymbol{y}) - \boldsymbol{y}^*(x+h)| \leqq |\boldsymbol{y}-\boldsymbol{y}^*| + h\left[|\boldsymbol{f}(x,\boldsymbol{y}) - \boldsymbol{f}\,(x,\boldsymbol{y}^*)| + \boldsymbol{\delta}(h)\right],$$

wobei $\boldsymbol{y} = \boldsymbol{y}^*(x)$ eine durch $(x^*, \boldsymbol{y}^*)$ gehende Integralkurve ist, die für $x^* \leqq x < x^* + \varepsilon$ in $\mathfrak{E}$ liegt, und $\delta(h)$ mit $h$ nach Null strebt. Also

$$\boldsymbol{\psi}(x+h, \boldsymbol{y}+h\boldsymbol{f}) - \boldsymbol{\psi}(x,\boldsymbol{y}) \leqq h\left[|\boldsymbol{f}(x,\boldsymbol{y}) - \boldsymbol{f}\,(x,\boldsymbol{y}^*)| + \delta(h)\right]$$

folglich nach (4) und (5), wobei $x = x^*$ ist,

$$\overline{\boldsymbol{D}}_{+[\boldsymbol{f}]}\boldsymbol{\psi}(x,\boldsymbol{y}) \leqq K\boldsymbol{\psi}(x,\boldsymbol{y})$$

Nun setzen wir $\varphi(x,\boldsymbol{y}) = \boldsymbol{\psi}(x,\boldsymbol{y})e^{-Kx}$, so ist $\varphi(x,\boldsymbol{y})$ von der Klasse $(L)$ und genügt in $\mathfrak{D}_1$ den Bedingungen: $\overline{\boldsymbol{D}}_{+[\boldsymbol{f}]}\varphi(x,\boldsymbol{y}) \leqq 0, \varphi(x,\boldsymbol{y}) = 0$ für $(x,\boldsymbol{y}) \in \mathfrak{E}_1$ und $\varphi(x,\boldsymbol{y}) > 0$ für $(x,\boldsymbol{y}) \in \mathfrak{D}_1 - \mathfrak{E}_1$. W.z.b.w. □

## §5. Vergleich eines Gleichungssystems mit einer einzigen Gleichung.

**Hilfsatz 2.** *Es sei $F(x,y)$ im Bereich $\mathfrak{D} : a \leqq x < b, y < \Lambda(x)$ stetig und sei der Bereich $\mathfrak{E} : a \leqq x < b, y \leqq \omega(x)$ in $\mathfrak{D}$ nach rechts majorant für*

$$\frac{dy}{dx} = F(x,y) \tag{6}$$

*wobei $\Lambda(x)$ und $\omega(x)$ in $a \leqq x < b$ stetig sind und $\omega(x) < \Lambda(x)$. Es gibt dann ein Unterintegral $\varphi(x,y)$ von (6) in $\mathfrak{D}_1 : a \leqq x \leqq b_1 < b$, $y \leqq \Lambda_1(x)$, wobei $\Lambda_1(x)$ in $\langle a,b)$ stetig und $\omega(x) < \Lambda_1(x) < \Lambda(x)$, derart, dass $\varphi(x,y) > 0$ ist für $y > \omega(x)$, $\varphi(x,y) = 0$ für $y \leqq \omega(x)$ und $\varphi(x,y)$ mit $y$ monoton wächst im erweiterten Sinne.*

*Beweis.* Es gibt eine Konstante $M > 0$, sodass in $\mathfrak{D}_1$ $|F(x,y)| < M$ ist.

Es sei $D(P,X)$ die Okamurasche Funktion von (6) in $\mathfrak{D}_1$, wobei $P = (\xi, \eta)$, $X = (x,y)$ und $\xi \leqq x$. Sind $\eta \leqq \omega(\xi)$, und $y \geqq \omega(x)$, so wächst $D(P,X)$ mit $y$ monoton im erweiterten Sinne.

Denn es seien $X = (x,y)$ und $\tilde{X} = (x,\tilde{y})$ Punkte mit demselben Wert von $x$, sodass $x > \xi$ und $\tilde{y} > y \geqq \omega(x)$. Für eine beliebige $\varepsilon > 0$ gibt es Punktfolge $\{P_i\}$ und $\{Q_i\}$ wie im Beweis von Satz 5 , sodass $P_0 = P$, $Q_n = \tilde{X}$ und

Es sei $\mathfrak{S}$ die Kurve, die aus Stücken $Q_iP_{i+1}$ der Integralkurven und vertikalen Strecken $P_iQ_i$ bestehen. Eine Integralkurve durch $X$ trifft $\mathfrak{S}$ an einem Punkt $Q^*$. $Q^*$ ist entweder auf einem Strecke $P_iQ_i$ oder auf einem Stücke $Q_iP_{i+1}$ der Integralkurve. Dann ist

$$D(P,X) \leqq \overline{P_0Q_0} + \cdots + \overline{P_iQ_i} < D(P,\tilde{X}) + \varepsilon$$

Folglich $D(P,X) \leqq D(P,\tilde{X})$.

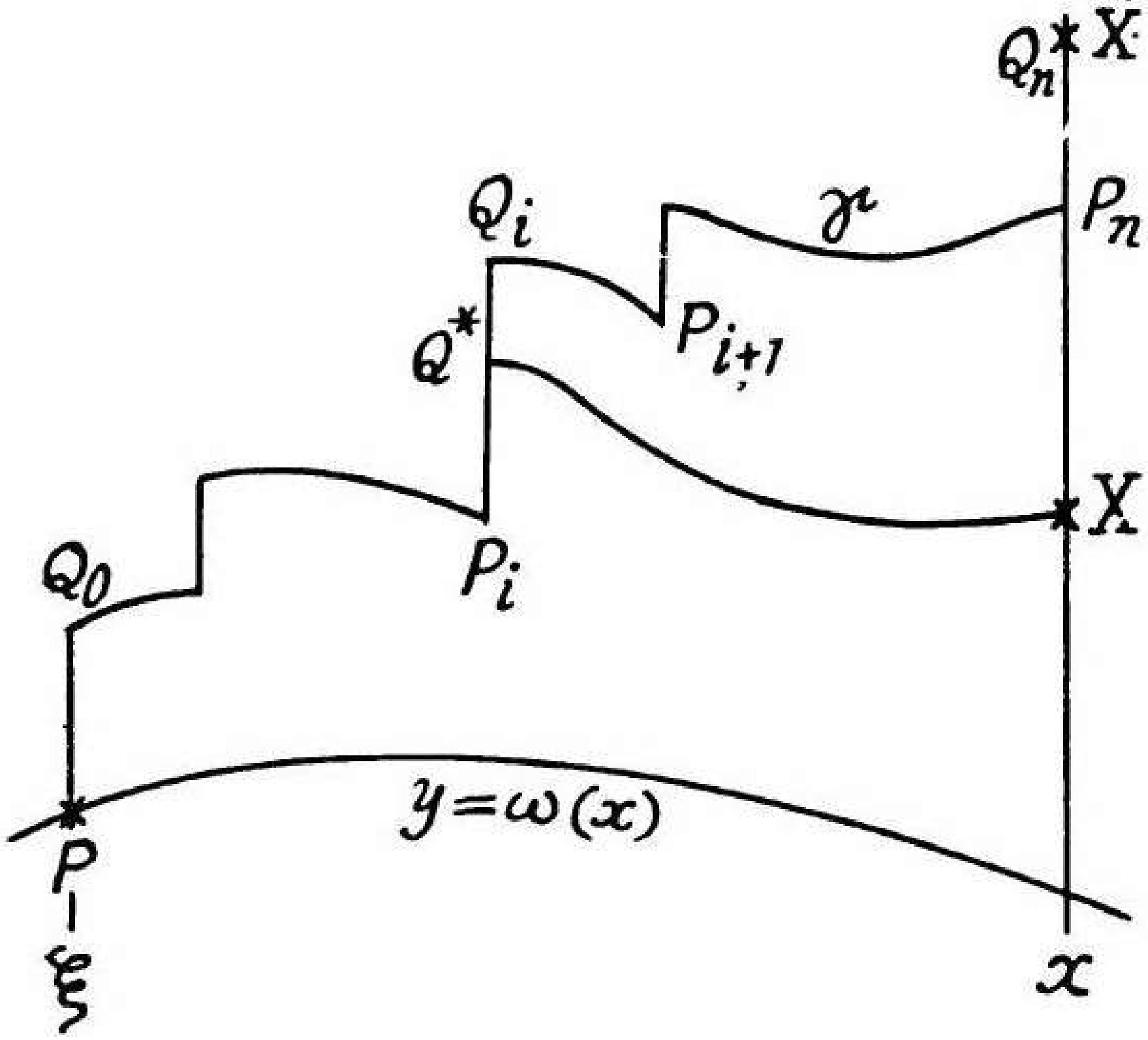

Fig. 2

$$\sum_{i=0}^{n} \overline{P_i Q_i} < D(P, \tilde{X}) + \varepsilon$$

Für $P \epsilon \mathfrak{E}_1$ und $X \in \mathfrak{D}_1 - \mathfrak{E}_1$ setzen wir

$$D^*(P, X) = \begin{cases} D(P, X), \text{ wenn } x \geqq \xi \\ y - \omega(x) + 2M(\xi - x), \text{ wenn } x < \xi. \end{cases}$$

Wir definieren dann $\varphi(X)$ durch

$$\varphi(X) = \begin{cases} \underset{P \in \mathfrak{E}_1}{\text{Min}}\, D^*(P, X), \text{ wenn } X \in \mathfrak{D}_1 - \mathfrak{E}_1 \\ 0, \text{ wenn } X \epsilon \mathfrak{E}_1. \end{cases}$$

So besitzt $\varphi(X)$ alle im Satz erwähnten Eigenschaften. W.z.b.w. □

**Satz 7.** *Es sei $F(x, y)$ im Bereich $a \leqq x < b$, $-\infty < y < \infty$ stetig und sei der Bereich $a \leqq x < b$, $y \leqq \omega(x)$ nach rechts majorant für die Gleichung*

$$\frac{dy}{dx} = F(x, y) \tag{6}$$

*wobei* $\omega(x)$ *in* $a \leqq x < b$ *stetig ist.*

*Nun sei* $\boldsymbol{f}(x, \boldsymbol{y})$ *im Bereich* $a \leqq x < b$, $|\boldsymbol{y}| < +\infty$ *stetig und* $S(x, \boldsymbol{y})$ *sei da von der Klasse* $(L)$ *mit der Eigenschaft:*

$$\overline{\boldsymbol{D}}_{+[\boldsymbol{f}]} S(x, \boldsymbol{y}) \leqq F(x, S(x, \boldsymbol{y})) \tag{7}$$

*Dann ist der durch* $a \leqq x < b$, $S(x, \boldsymbol{y}) \leqq \omega(x)$ *definierte Bereich* $\mathfrak{E}$ *nach rechts majorant für die Gleichung*

$$\frac{d\boldsymbol{y}}{dx} = \boldsymbol{f}(x, \boldsymbol{y}) \tag{1}$$

*Beweis.* Nach Hilfssatz 2 gibt es eine Funktion $\varphi(x, y)$ von der Klasses $(L)$ im $a \leqq x \leqq b_1 < b$, $|y| \leqq \Lambda$, sodass $\varphi(x, y) > 0$ für $y > \omega(x)$ und $\varphi(x, y) = 0$ für $y \leqq \omega(x)$,

$$\overline{\boldsymbol{D}}_{+[\boldsymbol{F}]} \varphi(x, y) \leqq 0 \tag{8}$$

und $\varphi(x, y)$ mit $y$ monoton wächst im erweiterten Sinne. Wir setzen $\Phi(x, \boldsymbol{y}) = \varphi(x, S(x, \boldsymbol{y}))$. $\Phi(z, \boldsymbol{y})$ gehört dann zur $(L)$. Nach (7) gilt für $h > 0$

$$S(x + h, \boldsymbol{y} + h\boldsymbol{f}) \leqq \boldsymbol{S}(x, \boldsymbol{y}) + h[F(x, S(x, \boldsymbol{y}) + \delta(h)]$$

wobei $\delta(h)$ mit $h$ nach Null strebt. Also

$$\Phi(x + h, \boldsymbol{y} + h\boldsymbol{f}) - \Phi(x, \boldsymbol{y}) \leqq \varphi(x + h, S + hF) - \varphi(x, S) + \alpha h\delta(h),$$

folglich nach (8)

$$\overline{\boldsymbol{D}}_{+[\boldsymbol{f}]} \Phi(x, \boldsymbol{y}) \leqq 0.$$

$\Phi(x, \boldsymbol{y})$ ist also ein Unterintegral von (1), $\Phi(x, \boldsymbol{y}) = 0$ für $(x, \boldsymbol{y}) \in \mathfrak{E}$ und $\Phi(x, \boldsymbol{y}) > 0$ für $(x, \boldsymbol{y}) \in \mathfrak{D} - \mathfrak{E}$. $\mathfrak{E}$ ist dann nach rechts majorant. W. z.b.w. ☐

Als ein spezialer Fall von 7 gilt:

**Satz 8.** *Es seien* $F(x, y)$ *und* $\boldsymbol{f}(x, \boldsymbol{y})$ *Funktionen, von denselben Eigenschaften wie in Satz 7, während die Ungleichung (7) durch*

$$S(\boldsymbol{f}(x, \boldsymbol{y})) \leqq F(x, S(\boldsymbol{y})) \tag{9}$$

*ersetzt ist, wobei* $S(\boldsymbol{y})$ *zur Klasse (L) gehört und genügt*

$$D_+ S(\boldsymbol{y}(x)) \leqq S(D_+ \boldsymbol{y}(x))$$

*für eine beliebige nach rechts differentiierbare Funktion* $\boldsymbol{y}(x)$, *z. B.* $\boldsymbol{S}(\boldsymbol{y}) = |\boldsymbol{y}|$.

*Ist der Bereich* $a \leqq x < b$, $y \leqq \omega(x)$ *nach rechts majorant für (6), wobei* $\boldsymbol{\omega}(x)$ *in* $a \leqq x < b$ *stetig ist, so ist der durch* $a \leqq x < b$, $S(\boldsymbol{y}) \leqq \boldsymbol{\omega}(x)$ *definierte Bereich nach rechts majorant für (1).*

Mathematisches Institut der Kaiserlichen Universität zu Osaka.